%% file: NJP_submission.tex
\documentclass[12pt]{iopart}

\usepackage{graphicx}
\usepackage{physics}
\usepackage{dsfont}
\usepackage[toc,page]{appendix}
\usepackage{braket}
\renewcommand\bra[1]{{\langle{#1}|}}
\renewcommand\ket[1]{{|{#1}\rangle}}
\usepackage{amsthm}
\usepackage{mathtools}
\usepackage{amssymb}
\usepackage{amsmath}
\usepackage{xcolor}
\usepackage[ocgcolorlinks,colorlinks=true,linkcolor=blue,citecolor=red]{hyperref}
\usepackage{verbatim}
\usepackage{hyperref}
\hypersetup{
	colorlinks=true,
	linkcolor=blue,
	filecolor=magenta,      
	urlcolor=cyan,
}
\usepackage{float}
\usepackage{rotating}
\usepackage{soul}
\input{commands.tex}
\usepackage{cite}

\makeatletter

\makeatletter

\begin{document}

\title{Finite-time dynamics of an entanglement engine: current, fluctuations and kinetic uncertainty relations}

\author{Jeanne Bourgeois$^1$, Gianmichele Blasi$^1$, Shishir Khandelwal$^{2,3}$, G\'eraldine Haack$^1$}
\address{$^1$Department of Applied Physics, University of Geneva, 1211 Geneva, Switzerland\\
$^{2}$Physics Department, Lund University, Box 118, 22100 Lund, Sweden\\
$^{3}$NanoLund, Lund University, Box 118, 22100 Lund, Sweden}	
\vspace{10pt}

\begin{abstract}
Entanglement engines are autonomous quantum thermal machines designed to generate entanglement from the presence of a particle current flowing through the device. In this work, we investigate the functioning of a two-qubit entanglement engine beyond the steady-state regime. Within a master equation approach, we derive the time-dependent state
, the particle current, as well as the associated current correlation functions. Our findings establish a direct connection between coherence and internal current, elucidating the existence of a critical current that serves as an indicator for entanglement in the steady state. We then apply our results to investigate kinetic uncertainty relations (KURs) at finite times. We demonstrate that there are more than one possible definitions for KURs at finite times. While the two definitions agree in the steady-state regime, they lead to different parameter's ranges for violating KUR at finite times.
\end{abstract}

\section{Introduction}

Quantum thermal machines have been proposed to perform a plethora of tasks, such as work production, refrigeration, metrology and time-keeping \cite{Kosloff2014,Vinjanampathy2016,Mitchison2019,Bhattacharya2021,Myers2022}. A particular class of thermal machines, \textit{entanglement engines}, has been shown to produce quantum entanglement autonomously, \textit{i.e.} by utilising only uncontrolled dissipation with thermal environments. These machines therefore produce a genuinely quantum output. The underlying mechanism for sustaining the presence of quantum coherence and entanglement in the steady-state regime from out-of-equilibrium environments is an established research direction for the last two decades \cite{Eisler2005, Hartmann2007, Quiroga2007, Bellomo2013, BohrBrask2015, Hegwill2018, Man2019, Farina23}. This series of works led to important and novel questions,  for example, the possibility of generating highly entangled and multipartite entangled states, the certification of entanglement from non-equilibrium transport observables and the utility of these machines for practical quantum information tasks. In recent years, key progresses have been achieved, leading to proposals for autonomous entanglement engines generating high-dimensional ~\cite{Tavakoli2018} and genuine multipartite entangled states~\cite{Tavakoli2020,Khandelwal2024}. 
These advancements have also motivated investigations of the minimal resources necessary for verifying the presence of entanglement through transport observables~\cite{Khandelwal2020}, as well as  for generating quantum states suitable for various quantum tasks~\cite{BohrBrask2022, Khandelwal2024}.

It has also been shown that the quantum statistics governing the reservoirs profoundly impact the amount of entanglement achievable in the steady state~\cite{BohrBrask2015, Prech2023, Khandelwal2024}. Experimental proposals involving semiconducting quantum dots~\cite{Das2022} and NV centers~\cite{Khandelwal2023} further enhance this understanding. As expected, the limitations on autonomous entanglement generation can be surpassed by harnessing additional resources, such as squeezed thermal baths~\cite{Tacchino2018}, non-Markovian environments~\cite{Heineken2021}, external drives~\cite{Aguilar2020, Khandelwal2023}, and feedback protocols~\cite{Diotallevi2023}. Notably, all these works predominantly focus on the steady-state regime of these entanglement engines.
\\

\par
In this article, we investigate the role of currents and current fluctuations in the functioning of an entanglement engine, specifically in the transient regime, \textit{i.e.}, for all times. Working within a Lindblad-equation approach, we obtain analytical results  for these quantities, exploiting recent results based on a quantum master equation and generalized full counting statistics (FCS) approach \cite{Blasi2023}. At finite times, we find that a minimum average current is not a sufficient condition for certifying the presence of entanglement. This is in contrast to the steady-state regime, in which a critical current was derived in Ref. \cite{Khandelwal2020} to witness entanglement, see also \cite{Heineken2021} 
for a related proposal. We then investigate the role of current correlation functions in this engine, through kinetic uncertainty relations (KURs) \cite{Garrahan2017,Terlizzi2019,Hiura2021}. These are classical relations that bound the signal-to-noise ratio (SNR), of particular interest for experimental applications, and for a fundamental understanding of quantum fluctuations in these engines.\\

The manuscript is organized as follows. In Sec.~\ref{sec:model}, we introduce the Hamiltonian model for a two-qubit entanglement engine and the Lindblad master equation that we solve for all times. We also recall the definitions of the current, the activity and the current correlation functions obtained within a FCS in \cite{Blasi2023}. 
In Sec.~\ref{sec:analytics}, we provide analytical expressions for the current and the coherence, at all times, for the two-qubit engine, for three different types of initial states. We discuss the dynamics of these quantities, together with the current correlation functions at all times. They allow us to discuss in Sec.~\ref{sec:witness} the validity of a critical current to certify the presence of entanglement beyond the steady-state regime. Finally, in Sec.~\ref{sec:KUR}, we introduce possible definitions of KUR at all times, investigating their violation as a function of times, temperature and voltage biases. We conclude with perspectives for future works.

\section{Model and definitions for a two-qubit entanglement engine}
\label{sec:model}

\vspace{0.2cm}

\subsection{Hamiltonian}

\begin{figure}[t]
\centering
\includegraphics[width=10 cm]{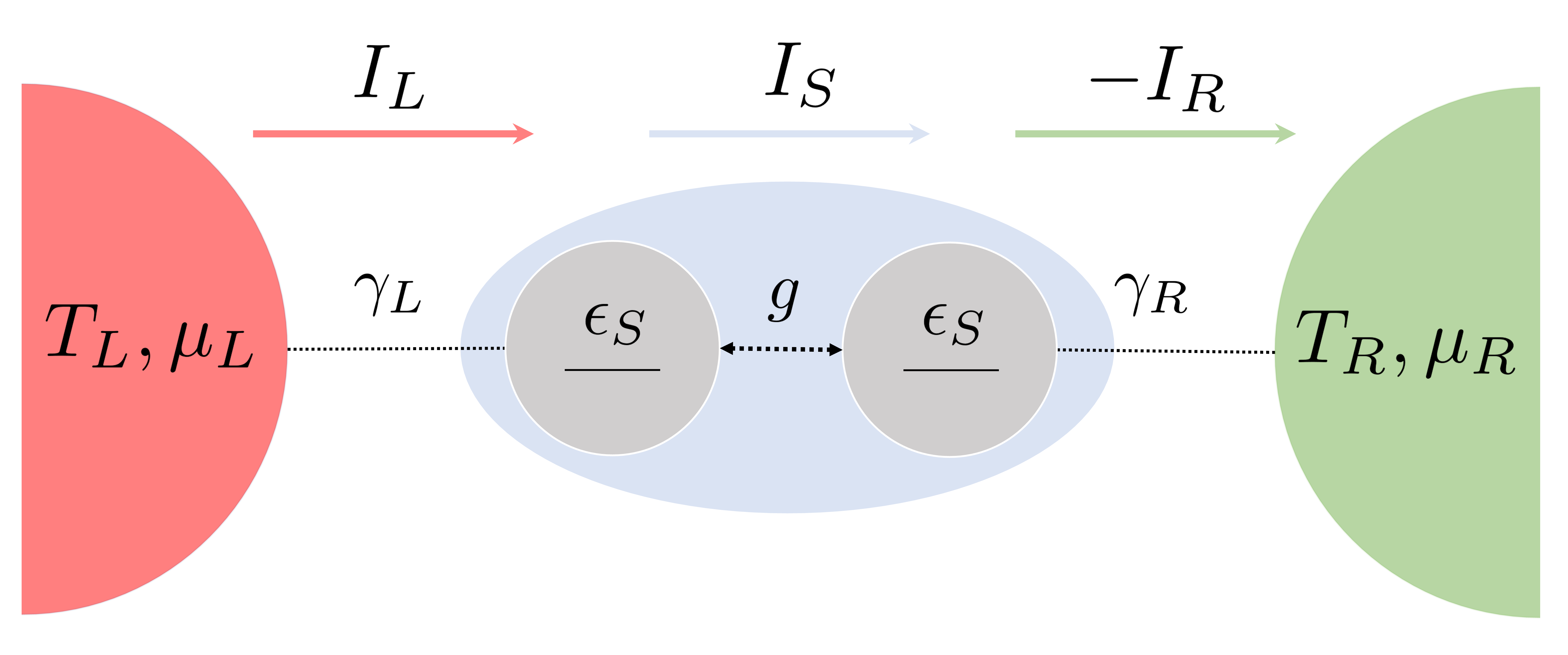}
 \caption{Sketch of the system. Two qubits with energies $\epsilon_S$ are tunnel-coupled with strength $g$. Each qubit is weakly coupled to a fermionic reservoir with temperature $T_{L/R}$ and chemical potential $\mu_{L/R}$. The bare coupling tunneling rates between the system and reservoir are denoted by $\gamma_{L/R}$. Arrows represent currents flowing between the left and right reservoir, \textit{i.e.}, $I_{L/R}$ respectively (Eqs.~\eqref{eq:current_all_times_1}), and $I_S$ denotes the internal current between the dots, see Eq.~\eqref{eq:IS}. Their relations at all times is discussed in, see Sec.~\ref{sec:witness}.} \label{fig:sketch}
\end{figure}

We consider a minimal model of two interacting qubits, labeled ``left" (L) and ``right" (R), independently coupled to two reservoirs at thermal equilibrium. The whole system evolves under the total Hamiltonian, $H = H_S + H_B + H_{SB}$, sum of 
\begin{itemize}
    \item the two-qubit Hamiltonian
\begin{equation}\label{eq:Hint}
    H_S := \epsilon_S ( \sigma_+^{(L)}\sigma_-^{(L)} + \sigma_+^{(R)}\sigma_-^{(R)} )+  H_{int},
\end{equation}
with $H_{int} := g(\sigma_+^{(L)}\sigma_-^{(R)}+\sigma_+^{(R)}\sigma_-^{(L)})$ the interaction Hamiltonian between the two qubits, $\sigma_{+}^{(j)} (\sigma_{-}^{(j)})$ the raising (lowering) operator for qubit $j$, $\epsilon_S$ the degenerate bare energy of both qubits and $g$ the flip-flop interaction strength;

    \item the two baths Hamiltonian
\begin{equation}
    H_B := \sum_{j=L,R} \sum_k \epsilon_{kj} c_{kj}^{\dag} c_{kj},
\end{equation}
with $c_{kj}^{\dag} ( c_{kj})$ the creation (annihilation) operator of the mode $k$, with energy $\epsilon_{kj}$, of bath $j$;

    \item and a system-bath tunneling Hamiltonian
\begin{equation}
    H_{SB} := \sum_{j=L,R}\sum_{k} (\alpha_{jk} \, \sigma_-^{(j)} \, c_{kj}^{\dag} + \alpha_{jk}^* \, c_{kj} \,\sigma_+^{(j)})
\end{equation}
    that describes the interaction between qubits and reservoirs, with $\alpha_{jk}$ the tunneling amplitude between the qubit $j$ and the $k$-th mode of bath $j$.
\end{itemize}
In the following, we set $\hbar=k_B=1$. While the above two-qubit model holds for both fermionic and bosonic baths, we focus on fermionic baths in this work, considering the anti-commutation relations for the bath operators $\{c_{kj},c_{k'j^\prime}^{\dagger}\}=\delta_{jj^\prime}\delta_{kk'}\I$. We also assume energy-degenerate qubits, $\epsilon_L = \epsilon_R := \epsilon_S$. The case of non-degenerate qubits in the steady-state regime was discussed in Ref.~\cite{Khandelwal2020,Potts2021}. A sketch of the setup is shown in Fig.~\ref{fig:sketch}.

\subsection{Lindblad master equation}

Under the assumption of a weak system-bath coupling regime($\gamma_j \ll \epsilon_S$), and a small interaction strength between the two qubits as compared to the couplings between them and their respective reservoirs ($g \lesssim \gamma_j$), the Markovian dynamics of the engine is well-captured by a so-called local Lindblad equation, which considers local transitions at the qubits induced by the jump operators $\sigma^{(j)}_\pm$ ~\cite{Breuer2007, Schaller2014,Hofer2017},
\begin{align} 
\label{eq:DQD_ME}
    \Dot{\densmat}(t) 
    &= \mathcal{L}\densmat(t) = - i [H_S, \densmat(t)] + \sum_{j=L, R} \left ( \gamma_j^+ \mathcal{D}[\sigma_+^{(j)}]\densmat(t) + \gamma_j^- \mathcal{D}[\sigma_-^{(j)}]\densmat(t) \right )\,.
\end{align}
Here, we use as notations for the excitation and de-excitation rates respectively $\gamma_j^+ = \gamma_j f_j(\epsilon_S)$ and $\gamma_j^- = \gamma_j (1-f_j(\epsilon_S))$. The bare rates $\gamma_j$ are set by the coupling strength $\alpha_{kj}$ in $H_{SB}$, and are assumed to be energy-independent in the wide-band limit~\cite{Baldea2016,Covito2018,Blasi2023}. The Fermi distribution of bath $j$ is defined as
\begin{equation}
    f_j(\epsilon) = \frac{1}{1+\e^{(\epsilon-\mu_j)/T_j}},
\end{equation}
with $T_j$ and $\mu_j$ its temperature and chemical potential. The rates $\gamma_j^-$ include spontaneous and stimulated emission processes due to the presence of bath $j$, while $\gamma_j^+$ include absorption due to reservoir $j$. The dissipators $\mathcal{D}[\mathcal{A}]\bullet = \mathcal{A}\bullet\mathcal{A}^{\dag} - \{\mathcal{A}^{\dag}\mathcal{A}, \bullet\}/2$ capture dissipation due to the presence of the reservoirs.
To discuss the activity in the context of KUR (see Sec.~\ref{sec:KUR}), it is convenient to decompose the Lindbladian super-operator $\mathcal{L}$ as~\cite{Schaller2014,Landi2023,Blasi2023} 
\begin{equation}
 \mathcal{L}=   \mathcal L_0 + \sum_{j=L,R  } (\mathcal L^+_{j} +\mathcal L^-_{j}),
\end{equation}
where
\begin{equation}
    \mathcal{L}_0 \bullet:= - i [H_S, \bullet] - \frac{1}{2} \sum_{j=L,R} \left ( \gamma_j^+ \{ \sigma_-^{(j)}\sigma_+^{(j)}, \bullet\} + \gamma_j^- \{ \sigma_+^{(j)}\sigma_-^{(j)}, \bullet \} \right )
\end{equation} 
accounts for coherent non-unitary evolution of the qubits, 
while
\begin{align}
\label{eq:jump_superop}
    \mathcal{L}_j^+ \bullet &:= \gamma_j^+ \sigma_+^{(j)} \bullet \sigma_-^{(j)}, \\
    \mathcal{L}_j^- \bullet &:= \gamma_j^- \sigma_-^{(j)} \bullet \sigma_+^{(j)},
\end{align}
represent quantum jumps, which can be seen as a result of continuous monitoring of the system by the environment \cite{Wiseman,Landi2023}, and cause transitions within the qubits.

\subsection{Lindblad equation: transient solution}

\vspace{0.2cm}

It is possible to solve the Lindblad equation~\eqref{eq:DQD_ME} 
through the vectorization of the density matrix. This amounts to recasting states and superoperators respectively as $16\times1$ vectors and $16\times16$ matrices, $\densmat(t) \longleftrightarrow \mathbf{p}(t),\; \; \mathcal{L} \longleftrightarrow \mathbf{L}$, and transforms the linear differential equation into the matrix differential equation, $\dot{\mathbf{p}}(t) = \mathbf{L}\,\mathbf{p}(t)$.
As noted in earlier works \cite{BohrBrask2015, Khandelwal2021}, the evolution as given by Eq.~\eqref{eq:DQD_ME} preserves the form of density matrices in the canonical basis $\{\ket{00}, \ket{01}, \ket{10}, \ket{11}\}$,
\begin{equation}
\label{eq:densmat}
    \densmat=
\begin{pmatrix}
r_1 & 0 & 0 & 0 \\
0 & r_2 & i c & 0 \\
0 & -i c^* & r_3 & 0 \\
0 & 0 & 0 & r_4 \\
\end{pmatrix}.
\end{equation}
This is
due to the form of the inter-qubit interaction Hamiltonian $H_{int}$, proportional to $\sigma_+^{(L)}\sigma_-^{(R)} +\sigma_+^{(R)}\sigma_-^{(L)}$. The elements $r_j$ are the populations of the canonical two-qubit states, satisfying $\sum_{j=1}^{4}r_j=1$. The element $c$ denotes the coherence from the $\ket{01}\bra{10}$ element of the density matrix. In particular, the steady-state solution of Eq.~\eqref{eq:DQD_ME} is of the form given by Eq.~\eqref{eq:densmat}, and its coherence $c^{(ss)}$ was shown to be real for degenerate qubits \cite{Khandelwal2020}.\\

In the next sections, we will discuss the presence of entanglement between the two qubits. This will be done by calculating the concurrence $\mathcal{C}$, a measure for bipartite entanglement, $0\leq\mathcal C\leq1$; $\mathcal C=0$ for separable states and $\mathcal C=1$ for maximally entangled states \cite{Wootters1998}. For density operators of the form of Eq.~\eqref{eq:densmat}, the concurrence can be simply calculated from the matrix elements, it takes the simple form \cite{Yu2007}, 
\begin{equation}
    \mathcal{C}(t) = \max \left \{ 0, 2(\abs{c(t)}-\sqrt{r_1(t)r_4(t)})\right \}. \\
\end{equation}

The density operator of Eq.~\eqref{eq:densmat} allows us to consider a reduced Lindbladian compared to Eq.~\eqref{eq:DQD_ME} that fully determines the evolution of the 6 elements of the density matrix. Its corresponding matrix is
\begin{equation}
\mathbf{L}_{red} = \left(
\begin{smallmatrix}
-(\gLp + \gRp)  & \gRm & \gLm & 0 & 0 & 0 \\
\gRp & -(\gLp + \gRm)  & 0 & \gLm & ig & -ig \\
\gLp & 0 & -(\gLm + \gRp)  & \gRm & -ig & ig \\
0 & \gLp & \gRp & -(\gLm + \gRm) & 0 & 0  \\
0 & ig & -ig & 0 & -\Gamma/2 & 0  \\
0 & -ig & ig & 0 & 0 & -\Gamma/2 
\end{smallmatrix}
\right)\,.
\end{equation}
The rates $\gamma_j^\pm$ have been defined below Eq.~\eqref{eq:DQD_ME}, and we have introduced the total bare rate $\Gamma = \gamma_L + \gamma_R$ for convenience. This matrix, written in the basis (in order) $\ket{00}\bra{00}$, $\ket{01}\bra{01}$, $\ket{10}\bra{10}$, $\ket{11}\bra{11}$, $\ket{01}\bra{10}$ and $\ket{10}\bra{01}$, is an operator acting on vectorized states in the Liouville space. The solution $\densmat(t)$ at all times is found by exponentiating the reduced Liouvillian, ${\mathbf{p}}(t) = e^{\mathbf{L}_{red}t} \, \mathbf{p}(0)$.
\\

\subsection{Average current and current correlation functions at finite times}

In this work, we investigate the currents and fluctuations of current exchanged between the baths and the two-qubit entanglement engine.
We derive them within a master equation approach, exploiting recent results from Ref.~ \cite{Blasi2023} and references therein. This allows us to express the average particle current from the reservoirs, $I_j(t)$, as well as the current correlation functions, $S_{jj'}(t,t')$, in terms of the corresponding current superoperator $\mathcal{I}_j$ and the dynamical activity superoperator $\mathcal{A}_j$,
\begin{equation}
\begin{aligned}
\label{eq:super_op}
    & \mathcal{I}_j:=\mathcal{L}_j^+ - \mathcal{L}_j^-\,, \\
    & \mathcal{A}_j:=\mathcal{L}_j^+ + \mathcal{L}_j^-\,,
    \end{aligned}
\end{equation}
with the jump superoperators $\mathcal{L}_j^{\pm}$ defined in Eqs.~\eqref{eq:jump_superop}. 
The expressions of the current and the correlation function take the following form
\begin{align}
\label{eq:current}
 I_j(t) = 
 \Tr{\mathcal{I}_j \densmat(t)}\,,  
\end{align}
and 
\begin{align} 
\label{eq:noise_1}
    S_{j j'}(t, t') &:= \delta_{j j'}\delta(t-t')\Tr{\mathcal{A}_{j}\densmat(t)} 
    + \Theta(t-t') \Tr{\mathcal{I}_{j}\e^{\mathcal{L}(t-t')}\mathcal{I}_{j'}\densmat(t')} \notag\\
    &\quad + \Theta(t'-t) \Tr{\mathcal{I}_{j'}\e^{\mathcal{L}(t'-t)}\mathcal{I}_{j}\densmat(t)}
    -\Tr{\mathcal{I}_{j}\densmat(t)} \Tr{\mathcal{I}_{j'}\rho(t')}\,,
\end{align}
with $\Theta$ the Heaviside function. We emphasize that the first term in the above expression only appears in the auto-correlation functions (due to the Kronecker delta $\delta_{jj'}$). It corresponds to the average dynamical  activity due to bath $j$, $A_j(t) := \Tr{\mathcal{A}_j \densmat(t)}$, which is a measure of the rate of jumps occurring at the interface with reservoir $j$, regardless of their direction. One can define the total activity $A(t)$ due to both left and right baths as
\begin{equation}
\label{eq:activity_tot}
    A(t) := A_L(t) + A_R(t)\,.
\end{equation}
A widely-discussed and relevant quantity is the shot-noise $\mathcal{S}$, defined as the zero-frequency component of the auto-correlation function in the limit of long times,
\begin{equation}
\mathcal{S}:=S_{LL}(\omega=0), \qquad S_{jj'}(\omega) :=\int_{-\infty}^{\infty} d\tau~ e^{-i \omega \tau} \lim_{t\to\infty} S_{jj'}(t, t+\tau).
\end{equation}
The steady-state correlation functions satisfy the relations
\begin{equation} \label{eq:corr_ss}
    S_{LL}(\omega=0)=S_{RR}(\omega=0)=-S_{LR}(\omega=0)=-S_{RL}(\omega=0),
\end{equation}
such that the shot-noise $\mathcal{S}$ describes equivalently the auto-correlations of the left and right currents, as well as the cross-correlations between these two currents, in the steady-state. \\
To compare the role of current correlation functions in an entanglement engine at finite and long times, one needs to define a single-time finite-frequency correlation function $S_{jj'}(\omega, t)$ that converges towards $S_{jj'}(\omega)$ for $t\to\infty$. In this work, we consider the following definition for $S_{jj'}(\omega, t)$,
\begin{equation}
    S_{jj'}(\omega, t) := \int_{-t}^t d\tau~ e^{-i \omega \tau} S_{jj'}(t, t+\tau).
\end{equation}
It corresponds to a pseudo-Fourier transform of the correlation functions $S_{jj'}(t,t')$ on the variable $t'$, symmetrised around $t'=t$. 
In particular, we will focus on its zero-frequency component,
\begin{equation}
\label{eq:noise}
   S_{jj'}(t) := S_{jj'}(\omega = 0, t) =  \int_{-t}^t d\tau~ S_{jj'}(t, t+\tau).
\end{equation}
The above definition indeed verifies $S_{jj'}(\omega, t)\to S_{jj'}(\omega)$ when $t\to\infty$. Furthermore, from Eq~\eqref{eq:noise}, by exchanging $t, t'$ and $j, j'$, one obtains the property,
\begin{equation}
S_{jj'}(\omega, t)=S_{j'j}(-\omega, t)\,.
\end{equation}
In particular, zero-frequency cross-correlations are equal at all times,
\begin{equation}
    S_{LR}(\omega=0, t)=S_{RL}(\omega=0, t).
\end{equation}

\section{Transient particle current, correlation functions and concurrence}
\label{sec:analytics}

\vspace{0.2cm}

In this section, we discuss the behavior of the current and correlation functions at all times. We consider three emblematic initial states for the two qubits: both qubits in their ground state denoted by $\densmat_0$, both qubits in a tensor product of thermal states denoted by $\densmat_{th}$, and both qubits being in a singlet state, a maximally entangled state, denoted $\densmat_{sg}$. The respective density operators are defined as:
\begin{align}
    &\densmat_0 = \ket{0}\bra{0}\otimes\ket{0}\bra{0}, \\
    &\densmat_{th} = \frac{\e^{-H_L/T_L}}{\Tr{\e^{-H_L/T_L}}}\otimes\frac{\e^{-H_R/T_R}}{\Tr{\e^{-H_R/T_R}}}, \\
    &\densmat_{sg} = \ket{\psi_+}\bra{\psi_+} = \frac{1}{2}(\ket{01}+\ket{10})(\bra{01}+\bra{01}).
\end{align}
We emphasize that all these initial states belong to the subset defined by Eq.~\eqref{eq:densmat}, hence they satisfy the condition for exploiting the reduced Lindbladian $\mathbf{L}_{red}$. We provide analytical expressions for the current and  the coherence for these three emblematic initial states and at all times. For each quantity, we clearly distinguish the time-independent terms, corresponding to the steady-state solutions, from exponentially decaying terms capturing the transient dynamics. \\

For the two qubits initially in their ground state $\densmat_0$, we obtain:
\begin{align}
    I_L(t)\lvert_{\densmat_0} =&
\frac{4g^2\gamma_L\gamma_R}{\Gamma(4g^2+\gamma_L\gamma_R)}(f_L-f_R) 
\notag \\
&+ \frac{\gamma_L}{\eta^2(4g^2+\gamma_L\gamma_R)}\text{e}^{-\frac{\Gamma}{2}t} 
\Big[ -\frac{16g^2}{\Gamma}(4g^2+\gamma_L\gamma_R)(\gL f_L + \gR f_R)
\notag \\
& \qquad +\cosh(\frac{\eta}{2}t)(4g^2\Gamma(\gL f_L + \gR f_R) + \gamma_L\gamma_R\eta^2 f_L) \notag \\
&\qquad +\eta\sinh(\frac{\eta}{2}t)(4g^2(-\gamma_L f_L + \gamma_R f_R) + \gamma_L\gamma_R (-\gamma_L+\gamma_R) f_L)
\Big] \,, \label{eq:current_rho0}\\
c(t)\lvert_{\densmat_0} =& - \frac{2g\gamma_L\gamma_R}{\Gamma(4g^2+\gamma_L\gamma_R)}(f_L-f_R) \notag \\
&+ \frac{g}{\eta^2}\text{e}^{-\frac{\Gamma}{2}t}
\Big[ 4\frac{\gamma_L-\gamma_R}{\Gamma}(\gL f_L+\gR f_R)  \notag \\
&\qquad +\frac{\Gamma}{4g^2+\gamma_L\gamma_R}\cosh(\frac{\eta}{2}t)\left((\gamma_R-\gamma_L)(\gL f_L+\gR f_R) + \frac{\eta^2}{\Gamma} (\gL f_L-\gR f_R)\right)  \notag \\
&\qquad + \frac{2 \gamma_L \gamma_R \eta}{4g^2+\gamma_L\gamma_R}\sinh(\frac{\eta}{2}t)(f_L-f_R) \Big]\,,
\end{align}
with $\Gamma = \gamma_L + \gamma_R$ and $\eta = \sqrt{(\gL-\gR)^2-16g^2}$. The latter factor determines the regime of the transient dynamics (overdamped, underdamped, critical, see \cite{Khandelwal2021} for the analysis of the non-Hermitian properties of this setup). The above expressions were obtained assuming $\eta>0 \in \mathbb{R}$, corresponding to the overdamped regime. For the two qubits initially in their thermal states $\densmat_{th}$, we obtain a similar structure for the current and coherence as a function of time: 
\begin{align}
    I_L(t)\lvert_{\densmat_{th}} =&
\frac{4g^2\gamma_L\gamma_R}{\Gamma(4g^2+\gamma_L\gamma_R)}(f_L-f_R) 
\notag \\
&+ \frac{4g^2\gamma_L}{\eta^2}(f_L-f_R) \text{e}^{-\frac{\Gamma}{2}t} 
\Big[ 2\frac{\gamma_R-\gamma_L}{\Gamma}
+ \frac{8g^2 - \gamma_R(\gamma_R-\gamma_L)}{4g^2+\gamma_L\gamma_R}\cosh(\frac{\eta}{2}t) \notag \\
& \qquad - \frac{\gamma_R}{4g^2+\gamma_L\gamma_R}\eta\sinh(\frac{\eta}{2}t)
\Big] \,, \label{eq:current_rhoth} \\
c(t)\lvert_{\rho_{th}} = 
&-\frac{2g\gamma_L\gamma_R}{\Gamma(4g^2+\gamma_L\gamma_R)}(f_L-f_R) \notag \\
&+\frac{2g\Gamma}{\eta^2}(f_L-f_R)\text{e}^{-\frac{\Gamma}{2}t}
\Big[\frac{(\gamma_L-\gamma_R)^2}{\Gamma^2} \notag \\
& \qquad -\frac{4g^2}{\Gamma(4g^2+\gamma_L\gamma_R)}(\Gamma\cosh(\frac{\eta}{2}t) + \eta \sinh(\frac{\eta}{2}t)) \Big] \,.\label{eq:coh_rhoth}
\end{align}
Finally, for the two qubits initially in the entangled state $\densmat_{sg}$, we obtain:
\begin{align}
    I_L(t)\lvert_{\densmat_{sg}} =&
\frac{4 \gamma_L \gamma_R g^2}{\Gamma (4 g^2 + \gamma_L \gamma_R)} (f_L - f_R) 
\notag \\
&+ \frac{\gamma_L}{4 (4 g^2 + \gamma_L \gamma_R) \eta^2} \text{e}^{-\frac{\Gamma}{2}t} 
\Big[32 \frac{g^2}{\Gamma} (4 g^2 + \gamma_L \gamma_R) (\gamma_L
(1 - 2 f_L) + \gamma_R (1 - 2 f_R)) \notag \\
& \qquad + 2 \gamma_L (-\gamma_L + \gamma_R) (1-2 f_L) \left( (\gamma_R (\gamma_L - \gamma_R) + 4 g^2) \cosh(\frac{\eta}{2}t) - \gamma_R \eta \sinh(\frac{\eta}{2}t) \right) \notag \\
& \qquad + 16 g^2 \gamma_L \gamma_R (-f_L + f_R) \cosh(\frac{\eta}{2}t) \notag \\
& \qquad + 8 g^2 (\gamma_L (1 - 2 f_L) - \gamma_R (1 - 
          2 f_R)) \left(\gamma_R \cosh(\frac{\eta}{2}t) + \eta \sinh(\frac{\eta}{2}t)\right)
\Big] \,,\\
c(t)\lvert_{\rho_{sg}} = 
& -\frac{4 g \gL \gR}{2 \Gamma (4 g^2 + \gL \gR)} (f_L - f_R) - \frac{i}{2}\text{e}^{-\frac{\Gamma}{2}t}  
\notag \\
& + \frac{g}{(4 g^2 + \gL \gR) \eta^2} \text{e}^{-\frac{\Gamma}{2}t}
\Big[ 2 \frac{-\gL + \gR}{\Gamma} (4 g^2 + \gL \gR)  \left( (1 - 2 f_L) \gL + (1 - 2 f_R) \gR \right) 
\notag \\
& \quad - 2 \cosh \left( \frac{\eta}{2}t \right) \left(4 g^2 (-(1 - 2 f_L) \gL + (1 - 2 f_R) \gR) + \gL \gR (-\gL + \gR) (1 - f_L - f_R) \right)
\notag \\
& \quad + 2 \sinh(\frac{\eta}{2}t) \gL \gR \eta (f_L - f_R)) 
\Big] \,.
\label{eq:coh_rhosg}
\end{align}
\par

\begin{figure}[t]
\begin{center}
\includegraphics[width=14 cm]{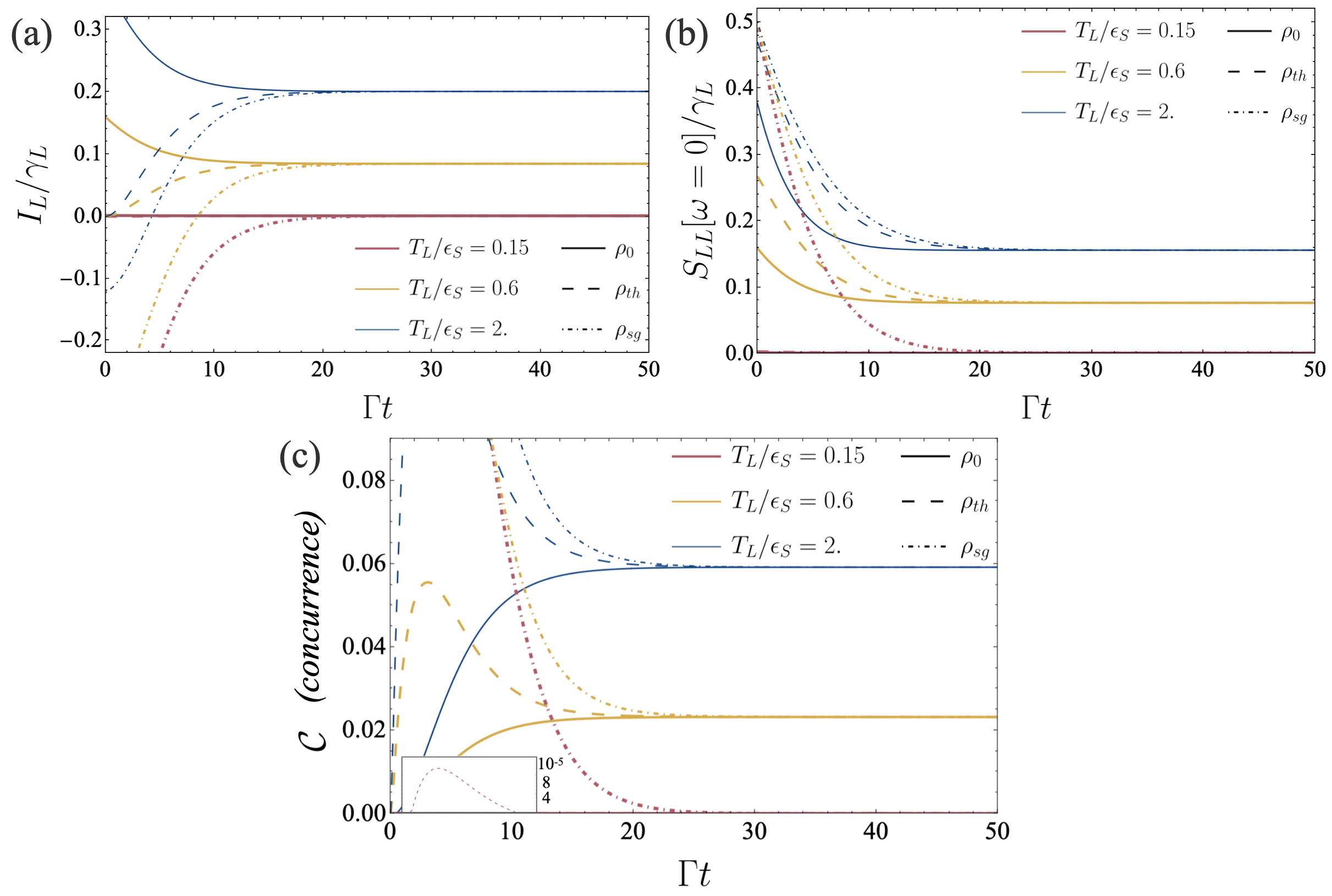}
\end{center}
\caption{Left current $I_L(t)$ (\textbf{a}), time-dependent zero-frequency left auto-correlations $S_{LL}(\omega=0,t)$ (\textbf{b}) and concurrence $\mathcal{C}(t)$ (\textbf{c}) as functions of time (in units of $\Gamma = \gamma_L + \gamma_R$). The current $I_L$ and its auto-correlation function $S_{LL}$ are normalised by their associated coupling rate, $\gL$. \\
Set of parameters: $\gamma_L/\epsilon_S=10^{-3}$, $\gamma_R/\epsilon_S=9\times10^{-3}$, $g/\epsilon_S=1.8\times10^{-3}$, $\mu_L/\epsilon_S=\mu_R/\epsilon_S=0$, $T_R/\epsilon_S = 0.1$, $T_L/\epsilon_S=0.15$ (red), $0.6$ (yellow) and $2$ (blue) and $\densmat(t=0) = \rho_0$ (solid), $\rho_{th}$ (dashed) and $\rho_{sg}$ (dot-dashed).\label{fig:time}}
\end{figure} 

These analytical results, together with the concurrence and zero-frequency current correlation functions are shown in Fig.~\ref{fig:time}. 
Here, we have fixed a constant cold temperature on the right, $T_R = 0.1$ (in units of $\epsilon_S$), while varying the temperature of the hot reservoir on the left. For the three temperature gradients, the current monotonically converges towards its steady-state value, see panel a). However, this monotonicity results from a specific choice of parameter, it does not constitute a general feature. A higher value of $g$ (precisely, $g> (\gR-\gL)/4$) would have led to oscillations in the transient regime, comparable to an under-damped evolution~\cite{Khandelwal2021}. For the two qubits initially in the singlet state, the current takes a negative value for short times. Analytics show that this originates in an imbalance between the mean occupation of the left reservoir set by $f_L$ and the initial occupation of the qubit. It is not specific to the presence of entanglement. The current correlations, panel b), similarly to the current, converge towards their steady-state values, which do not depend on the initial state. From our analytical expressions and the plot, we cannot identify specific features which could be attributed to the presence of entanglement. Finally, the concurrence exhibits a highly non-monotonous behaviour, see panel c). The steady-state value increases with the temperature bias, as expected. In the transient regime, higher values for $\mathcal{C}$ can be obtained, see for instance the case of the two qubits initially in a thermal state (red dashed curve). We also observe a decrease of the concurrence as a function of time for the qubits initially in a singlet state, a consequence of dissipation in the entanglement engine.\\

Let us comment on the validity of these solutions and behaviors at short times, close to $t_0=0$. It is important to keep in mind that these results were derived within a FCS approach, valid in the weak system-bath coupling regime. At times very close to 0, the current and current correlation functions take a finite value for generic initial states, while they are all 0 at $t=0$. This discontinuity directly stems from the wide-band limit approximation, which assumes energy-independent bare coupling rates $\gamma_{L,R}$ at all times, see discussions in Refs.~\cite{Covito2018,Blasi2023}. It is therefore crucial to emphasize that results obtained within a FCS approach hold significance only if the time interval $t-t_0$ is sufficiently large to compensate for the small value of the coupling rate imposed by the weak-coupling regime.

\section{Average current to certify the presence of entanglement}
\label{sec:witness}

\vspace{0.2cm}

\begin{figure}[H]
\begin{center}
\includegraphics[width=14 cm]{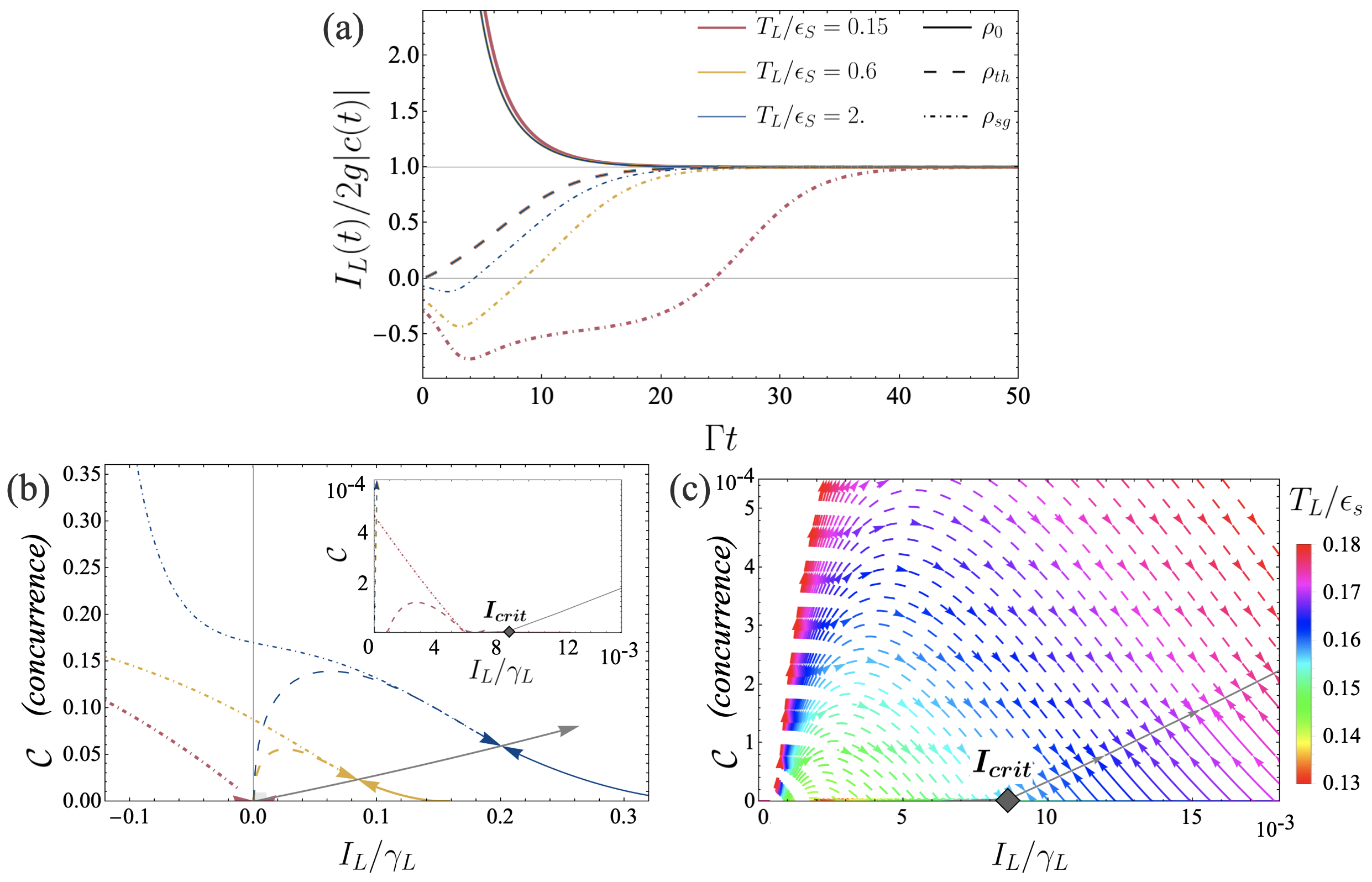}
\end{center}
\caption{Relation between current and creation of entanglement. (\textbf{a}) Ratio of current and coherence over time (in units of $\Gamma$). At long times, we recover the proportionality relation as found in \cite{Khandelwal2020}, for all temperature gradients and all initial states. Depending on the initial state of the qubits, the current can start from a negative value, see dotted-dashed lines for the qubits being in a singlet state at time $t=0$. (\textbf{b}) Parametric plot of the concurrence and the current, both as a function of time, $t$ varying from $0$ to infinity. For each temperature gradient (curves in a given color), the qubits end up to the same steady state, with a well-defined value for the concurrence $\mathcal{C}^{(ss)}$ and current $I_L^{(ss)}$, lying exactly on the grey curve. The intersection of the grey curve with the x-axis corresponds to the critical current, necessary to operate the engine successfully. (\textbf{c}). Zoom of the parametric plot shown in panel b) in the region of small currents. Different colored curves correspond to different temperature gradients (by varying $T_L$, see color grid). It becomes evident that for steady-state currents smaller than $I_\text{crit}$, concurrence is 0, while for steady-state currents larger than $I_\text{crit}$, the qubits are characterized by a finite value of $\mathcal{C}$. Set of fixed parameters: $\gamma_L/\epsilon_S=10^{-3}$, $\gamma_R/\epsilon_S=9\times10^{-3}$, $g/\epsilon_S=1.8\times10^{-3}$, $\mu_L/\epsilon_S=\mu_R/\epsilon_S=0$, $T_R/\epsilon_S = 0.1$. \label{fig:entanglement}}
\end{figure}

Recently, Ref.~\cite{Khandelwal2020} put forward the exact relation between a finite steady-state current and a finite value for the coherence, establishing for the first time the possibility to certify the presence of entanglement from an observable accessible in transport experiments. Explicitly, this relation states that the current is set by the coherence $c$ and inter-qubit interaction strength $g$, an exact result from deriving the currents in the reservoirs in the steady state. This relation can be seen from the exact expressions of the current and coherence at all times, see Eqs.~\eqref{eq:current_rho0}-\eqref{eq:coh_rhoth}. Indeed, in the steady-state (time-independent terms), it is straightforward to verify this close connection between coherence and current. This is shown in Fig.~\ref{fig:entanglement} a), where we plot the ratio $I_L/(2g\vert c \vert)$ as a function of time. At long times, it converges to 1, independently of temperature bias and initial state. \\

This behavior is in contrast to the transient regime, see Fig.~\ref{fig:entanglement} a), where this proportional relation between current and coherence clearly breaks down, for all temperatures, and all initial states. It is interesting to note a peculiar behavior when the two qubits are initially in a thermal state (dashed lines). In this case, the relation between current and coherence is independent of the temperature bias, at all times. We understand this result as a consequence of the qubits being initially populated according to the mean occupation of the bath at a given temperature. Given this initial condition, the dynamics of the qubits is fixed, leading to identical time evolution for all temperature gradients. This understanding is supported by our analytical results, see Eqs.~\eqref{eq:current_rhoth}-\eqref{eq:coh_rhoth}. Both the current and the coherence are proportional to $f_L-f_R$, such that their ratio becomes independent of these distributions, in particular independent of the temperatures and chemical potentials.\\


This relation between particle current and coherence can be explained by delving into the current conservation equations for the two-qubit system and the left and right baths. For the two-qubit system, the change of occupation probability of the left and right qubits can be defined respectively from their number operators $\hat{n}_j = \sigma_+^{(j)}\sigma_-^{(j)}, j=L,R$ as:
\begin{align}
&\Dot{n}_L = \Tr(\hat{n}_L \, \dot{\densmat}) =  \Tr(\sigma_+^{(L)}\sigma_-^{(L)} \mathcal{L}\densmat)\,, \\
&\Dot{n}_R = \Tr(\hat{n}_R \, \dot{\densmat}) =  \Tr(\sigma_+^{(R)}\sigma_-^{(R)} \mathcal{L}\densmat) \,,
\end{align}
with $\mathcal{L}$ the Lindbladian superoperator defined in Eq.~\eqref{eq:DQD_ME}. Distinguishing the unitary and dissipative parts, one obtains
\begin{align}
\label{eq:change_n_j}
    \dot{n}_j = - i\Tr(\sigma_+^{(j)}\sigma_-^{(j)}[H_S, \rho(t)]) + \Tr{ \sigma_+^{(j)}\sigma_-^{(j)} \sum_{s=\pm}  \gamma_j^s \mathcal{D}[\sigma_s^{(j)}]\densmat(t)}, \quad j=L,R\,. 
\end{align}
Due to particle number conservation within the system, $[n_L + n_R, H_S]=0$, we have 
\begin{align}
    -i\Tr(\sigma_+^{(L)}\sigma_-^{(L)}[H_S, \rho(t)]) = i\Tr(\sigma_+^{(R)}\sigma_-^{(R)}[H_S, \rho(t)])\,,
\end{align}
which we interpret as an internal current $I_S$ flowing through the system. Because $[n_j, n_L+n_R]=0, j=L,R$, this internal current only depends on $H_{int}$,
\begin{align}
\label{eq:IS}
 I_S(t) := i\Tr(\sigma_+^{(L)}\sigma_-^{(L)}[H_{int}, \rho(t)])\,.
\end{align}
Interestingly, for a density matrix of the form given by Eq.~\eqref{eq:densmat} and for $H_{int}$ given by Eq.~\eqref{eq:Hint}, the internal current takes the simple expression
\begin{equation}
    I_S(t) = -2g\Re\{c\}(t).
\end{equation}
In addition, due to total conservation of particles in the system and reservoirs, $\dot{n}_L + \dot{n}_R = I_L + I_R$, we can identify the dissipative parts in Eq.~\eqref{eq:change_n_j} as the particle current $I_j$ flowing into bath $j$ at time $t$, 
\begin{align}
    I_j(t) = \Tr{ \sigma_+^{(j)}\sigma_-^{(j)} \sum_{s=\pm}  \gamma_j^s \mathcal{D}[\sigma_s^{(j)}]\densmat(t)}\,.
\end{align}
We adopt the convention for $I_S$ to be positive when flowing from left to right qubits. To summarize, at all times, we can write
\begin{equation}
\begin{aligned}
    & \dot{n}_L(t) = - I_S(t) + I_L(t) \,, \label{eq:current_all_times_1} \\
    & \dot{n}_R(t) = I_S(t) + I_R(t) \,. 
    \end{aligned}
\end{equation}
In the steady state, we have $\dot{n}_L = \dot{n}_R=0$, leading to
\begin{align}
    I_L^{(ss)} = I_S^{(ss)} = - I_R^{(ss)}\,.
\end{align}
For a temperature/potential gradient such that particle currents are flowing from the left to the right bath, we thus obtain the relation $I_L^{(ss)} = 2g\abs{c^{(ss)}}$, see Eq.~\eqref{eq:IS}. In the transient regime, this does not hold. The more general equations \eqref{eq:current_all_times_1} must be considered.\\

In Fig.~\ref{fig:entanglement} panels b) and c), we further illustrate the tight relation between entanglement and current with parametric plots of the concurrence $\mathcal{C}$ and current $I_L$, both as functions of time. Independently of their initial state, the qubits converge towards their steady state, which differs depending on the temperature gradient. Blue, orange and red curves in panel a) correspond to different values of $T_L$ for fixed value of $T_R$. The grey curve corresponds to a parametric plot of the same quantities in the steady state, \textit{i.e.} it is a parametric plot of $\mathcal{C}^{(ss)}$ and $I_L^{(ss)}$, varying the temperature $T_L$ for fixed $T_R$. All colored curves for different $T_L$ and different initial states converge to this grey curve. Arrows on the colorful curves indicate the direction of the dynamics as a function of time. This parametric plot highlights the significance of the critical current $I_\text{crit}$. If $I_L$ takes a value smaller than $I_\text{crit}$ in the steady state, the corresponding concurrence is 0. This is clearly visible in panel c) (zoom of panel b) on small currents $I_L$).\\

\section{Kinetic uncertainty relations in a two-qubit entanglement engine}
\label{sec:KUR}

\vspace{0.2cm}

\subsection{Definitions of KURs in the transient regime}

In the context of classical Markovian systems in contact with one or several environments, Kinetic Uncertainty Relation (KUR) was introduced to set a bound to the signal-to-noise ratio of an observable $\mathcal{O}$, $\frac{ \langle\!\langle \mathcal{O}^2 \rangle\!\rangle}{\expval{\mathcal{O}}^2}$ (where $ \langle\!\langle \bullet \rangle\!\rangle$ represents the cumulant), in terms of the system's dynamical activity $A$ \cite{Garrahan2017,Terlizzi2019,Hiura2021}.
Within the domain of transport phenomena, KUR has been mainly discussed within the steady-state regime \cite{Timpanaro2019, Hiura2021, Kewming2023, Prech2023, Timpanaro2023}, akin to the exploration of thermodynamic uncertainty relations in quantum transport setups, as discussed in Refs.~\cite{Liu2021, Gerry2022, Ferri2023}.
Expressed formally, KUR takes the following form,
\begin{align}
\label{eq:KUR}
   \mathcal{R}^{(ss)} = \frac{S^{(ss)}}{(I^{(ss)})^2} A^{(ss)} \leq 1\,. 
\end{align}
At finite times, KURs have not been yet studied to the best of our knowledge. 
The study of time-dependent current-related KURs introduces several pertinent questions.
 Should we account for the right or left currents, as the two of them are not anymore equivalent in the transient regime? Or should we consider a combination of these two, for example symmetric or asymmetric with respect to the system? Note that for each chosen version of the current operator, its associated correlation function would be different, with auto- and cross-correlations being involved or not. This motivates our proposition  of two different KURs, defined via the ratios $R_L$ and $R_{asym}$ below,
\begin{eqnarray}
R_L(t) &:=&  \frac{S_{LL}(t)}{I_L(t)^2} \, A(t), \label{eq:KUR_L} \\
R_{asym}(t)&:=& \frac{S_{LL}(t)+S_{RR}(t)- S_{LR}(t) - S_{RL}(t)}{(I_L(t)-I_R(t))^2} \, A(t)\,. \label{eq:KUR_asym}   
\end{eqnarray}
Current, (zero-frequency) current correlation functions and activity were defined in Sec.~\ref{sec:model}, Eqs.~\eqref{eq:current}, \eqref{eq:noise} and \eqref{eq:activity_tot} respectively.\\

In the definition for $R_L$, Eq.~\eqref{eq:KUR_L}, we only consider the signal-to-noise ratio restricted to the left bath, set by $S_{LL}$ and $I_L$. In contrast, for $R_{asym}$, Eq.~\eqref{eq:KUR_asym}, we propose to bound the signal-to-noise ratio accounting for the average flow from the left to the right bath, $(I_L-I_R)/2$, with the associated noise $(S_{LL}-S_{LR}- S_{RL}+S_{RR})/4$. In both expressions, we propose to bound the SNR with the activity $A = A_L + A_R$, as defined in Eq.~\eqref{eq:activity_tot}. When considering quantum coherent transport through a multipartite system, the activity can in general also account for internal jumps occurring within the system. This additional contribution was first discussed in \cite{Prech2023}, and may lead to different conclusions concerning the possible violation of KURs. In our treatment, we adopt a \textit{mesoscopic} point of view, in which the internal dynamics of the quantum system is not accessible, only jumps between the system and baths are assumed to be accessible via the measurements of observables. We believe it is an open question to determine what is a valid bound in terms of the dynamical activity for coherent nanoscale devices, and we leave it open for future works. In the Appendix, we discuss KURs violation based on the ratios $R_L$ and $R_{asym}$ in Eqs.~\eqref{eq:KUR_L}-\eqref{eq:KUR_asym}, accounting for the additional internal activity. In agreement with \cite{Prech2023}, we find that this contribution prevents the violation of KURs in the steady state, for the set of parameters we considered. However, at finite times, KUR violation still occurs over a certain time range.\\


Importantly, in the long-time limit, the definitions Eqs.~\eqref{eq:KUR_L} and \eqref{eq:KUR_asym} coincide with each other, as well as with the \textit{standard} definition Eq.~\eqref{eq:KUR}, due to current conservation and steady-state relations~\eqref{eq:corr_ss},
\begin{equation}
    R_L^{(ss)} = R_{asym}^{(ss)} = \mathcal{R}^{(ss)}\,.
\end{equation}
Note that other definitions could have been chosen for the single-time finite-frequency correlation functions $S_{jj'}(\omega, t)$~\cite{Joho2012}. However, these definitions would only differ at finite times during the transient regime: at initial and long times, they all reduce to the same expressions. We also emphasize that we restrict this work to the investigation of a bound for the SNR. Tighter bounds, emphasizing quantum contributions, have been discussed in the context of Cram\'er-Rao bound with the quantum Fisher information \cite{VanVu2022}.


\subsection{Violation of KURs}

In this section, we investigate the behaviors of the ratios $R_L$ and $R_{asym}$ as functions of time, for different initial states, different temperature gradients and different potential imbalances. Let us first provide some analytical insights of the behavior of KUR ratios at very short and long times. Initially, KUR ratios equal to
\begin{align}
\label{eq:KUR_t_0}
    R_L(t=0) 
    = \frac{A_L(0)}{I_L(0)^2}(A_L(0)+A_R(0)), 
    \end{align}
    \begin{align}
    R_{asym}(t=0) 
    &= \frac{(A_L(0)+A_R(0))^2}{(I_L(0)-I_R(0))^2}.
\end{align}
Noting that the activity $A_j$ is always greater than the absolute value of the current $\abs{I_J}$, as
\begin{equation}
    A_L = \Tr(\mathcal{L}_j^+\densmat) + \Tr(\mathcal{L}_j^-\densmat), \qquad \text{with} \; \Tr(\mathcal{L}_j^{\pm}\densmat)\geq 0,
\end{equation}
and 
\begin{equation}
    \abs{I_L} = \abs{\Tr(\mathcal{L}_j^+\densmat) - \Tr(\mathcal{L}_j^-\densmat)} \leq \abs{\Tr(\mathcal{L}_j^+\densmat)} + \abs{\Tr(\mathcal{L}_j^-\densmat)} = A_j\,,
\end{equation}
we deduce that both KURs are always satisfied at time $t=0$. In the steady-state, we use the expressions for the current average and fluctuations, see Eq.~\eqref{eq:current_rho0} at $t\rightarrow \infty$ and \cite{Prech2023},
\begin{align}
    I^{(ss)} &= \frac{4g^2\gL\gR}{\Gamma(4g^2+\gL\gR)}(f_L-f_R), \\
    S_{LL}^{(ss)} &= I^{(ss)} \frac{f_L(1-f_R)+f_R(1-f_L)}{f_L-f_R} - 2 I^{(ss)2}\left (\frac{1}{\Gamma} + \frac{\Gamma}{4g^2 + \gL \gR} \right ),
\end{align}
and compute the steady-state activity,
\begin{align}
    A^{(ss)}= \frac{2\Gamma}{4g^2+\gL\gR}\left [(4g^2+\gL\gR) \avg{f(1-f)} + 4g^2\frac{\gL\gR}{\Gamma^2}(f_L-f_R)^2  \right ],
\end{align}
with the notation $\avg{x}=\frac{\gL x_L + \gR x_R}{\gL + \gR}$, to determine the full expression of the KUR ratios,
\begin{multline}
\label{eq:KUR_ss}
    R_L^{(ss)} = R_{asym}^{(ss)} = 
    4 \left [(4g^2+\gL\gR) \avg{f(1-f)} + 4g^2\frac{\gL\gR}{\Gamma^2}(f_L-f_R)^2 \right ] \\
    \times \left [\frac{\Gamma^2}{8g^2\gL\gR}\left (\frac{f_L(1-f_L)+f_R(1-f_R)}{(f_L-f_R)^2} + 1 \right ) - \frac{4g^2 + \gL \gR + \Gamma^2}{(4g^2 + \gL\gR)^2}\right ]\;.
\end{multline}
In the steady state, KUR violation depends on multiples parameters, making it difficult to assess a priori. However, we see that a greater imbalance $\abs{f_L - f_R}$ between the baths is necessary to reduce the KUR ratios, whereas greater temperatures $T_L$ and $T_R$ strictly increase them through the terms of form $f_j(1-f_j)$. This observation testifies for the need of a large enough energy bias $\abs{\mu_L - \mu_R}$ between the reservoirs, and low temperatures $T_L$ and $T_R$, to violate KUR. \\

\begin{figure}[H]
\begin{center}
    \includegraphics[width=14 cm]{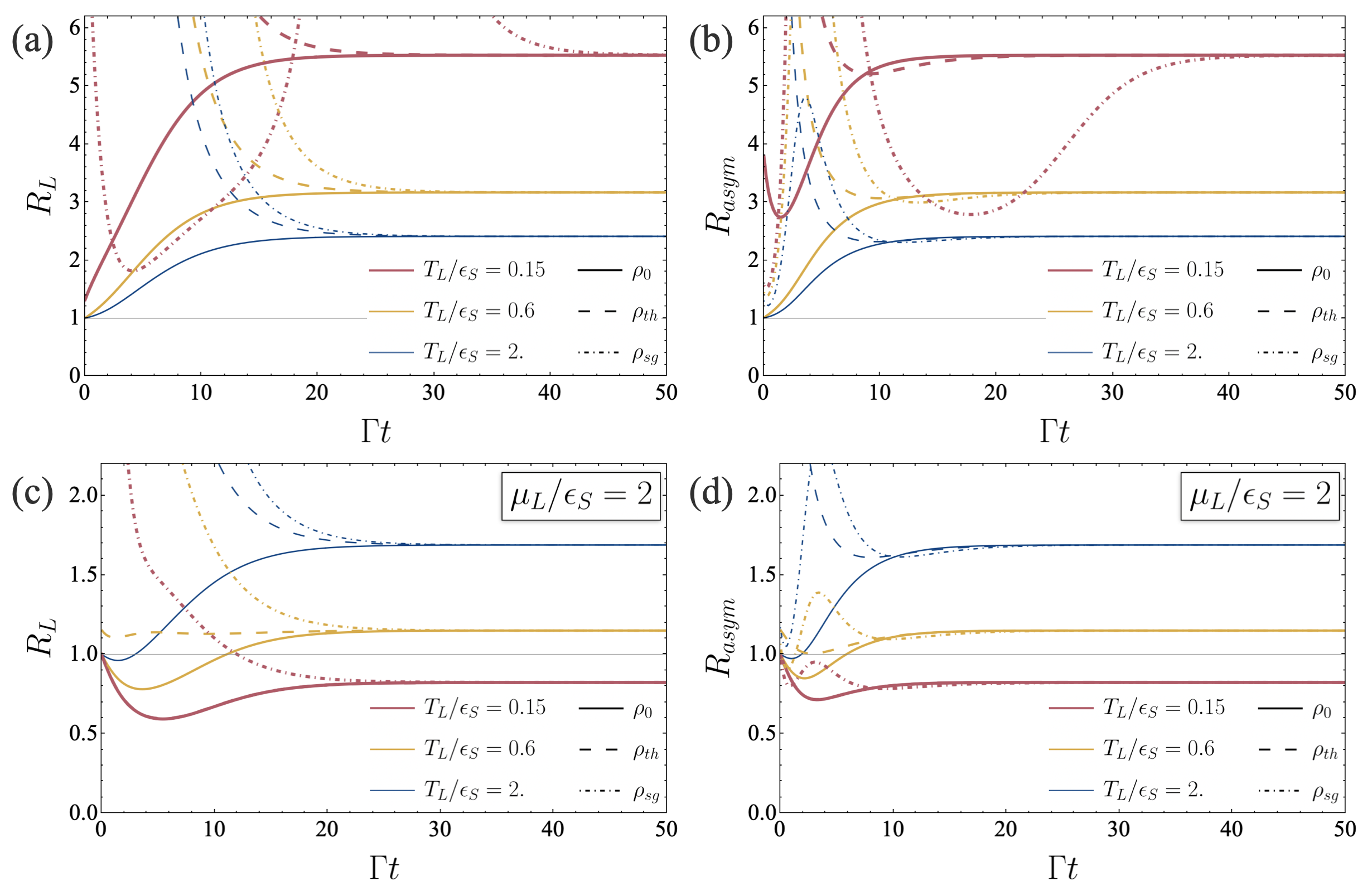}
\end{center}
\caption{``Left" (\textbf{a}, \textbf{c}) and ``asymmetric" (\textbf{b}, \textbf{d}) KUR ratios $R_L$ and $R_{asym}$ as functions of time, for $\mu_L/\epsilon_S=0$ (\textbf{a}, \textbf{b}) and $\mu_L/\epsilon_S=2$ (\textbf{c}, \textbf{d}). Set of parameters: $\gamma_L/\epsilon_S=10^{-3}$, $\gamma_R/\epsilon_S=9\times10^{-3}$, $g/\epsilon_S=1.8\times10^{-3}$, $\mu_R/\epsilon_S=0$, $T_R/\epsilon_S = 0.1$.\label{fig:KUR}}
\end{figure}

Figure \ref{fig:KUR} shows $R_L$ and $R_{asym}$ at all times. Panels a) and b) displays the KUR ratios for a parameter set identical as in Figs.~\ref{fig:time} and ~\ref{fig:entanglement}, in absence of a potential bias between the two reservoirs. In this case, despite a temperature bias, we never observe a violation of KURs over time, for all initial states we considered. In contrast, in presence of a energy potential bias, $eV=\mu_L-\mu_R=2\epsilon_S$, Fig.~\ref{fig:KUR} c) and d), KURs are violated over a certain time range, and even in the steady state at low temperature. This result supports the discussion below Eq.~\eqref{eq:KUR_ss}.



In the presence of a finite energy potential bias $eV$, the behaviors of KURs also differ at long times. In the steady-state, the dependence of KUR ratios to temperature gradient is opposite, small temperature gradients corresponding to steady-state violation of KURs. 
Violation at finite times highly depends on the initial state. Interestingly, out of the three considered initial states, the ground state provides the highest finite-time KUR violation, whereas the singlet state does not necessarily provide a smaller value for $R_L$ and $R_{asym}$ compared to the thermal state. We thus do not observe an enhancement of KUR violation due to the presence of entanglement at finite time. 

\section{Conclusions}

This work provides for the first time the analysis of the functioning of an entanglement engine at finite times. Within a master equation approach, assuming a weak inter-qubit coupling strength with respect to system-bath couplings, we provide analytical expressions for the particle current, the quantum coherence and the zero-frequency current correlation functions at finite times, utilising recent results \cite{Blasi2023,Landi2023}. 
Our results allow us to certify that a critical average current is not sufficient to certify the presence of entanglement beyond the steady-state regime. We also put in evidence two interesting signal-to-noise ratios to be bounded by the dynamical activity in the transient regime, taking into account transport observables in one or two reservoirs. Their transient behaviors do not differ qualitatively, only quantitatively, their steady-state values being the same. This result seems to indicate that these two KUR ratios capture the same fundamental physics and our analytical expressions indicate that a potential bias between the two reservoirs may be a better resource to violate KUR than a temperature bias.

\section{Acknowledgements}
G.H and G.B. acknowledge funding from the NCCR SwissMAP. S.K. acknowledges support from the Knut and Alice Wallenberg Foundation through the Wallenberg Center for Quantum Technology (WACQT).

\section*{References}

\bibliography{biblio}

\section*{Appendix}
\label{sec:Appendix}

In this Appendix, we discuss the effect of the internal activity $A_S$, the activity's contribution originating from internal jumps within the system, on the KUR. Its value at all times can be calculated from its definition in terms of the Liouvillian matrix, provided in \cite{Prech2023}:
\begin{equation}
    A_S(t) := \frac{4g^2}{\Gamma}(r_2(t)+r_3(t)).
\end{equation}
In the steady state, this internal activity takes the explicit form:
\begin{align}
    A_S^{(ss)}
    =& \frac{4 g^2}{\Gamma(4g^2+\gL\gR)} \Big[8g^2\avg{f}(1-\avg{f})+\gamma_L\gamma_R(f_L(1-f_R)+f_R(1-f_L))  \Big].
\end{align}

\begin{figure}[H]
\begin{center}
    \includegraphics[width=14 cm]{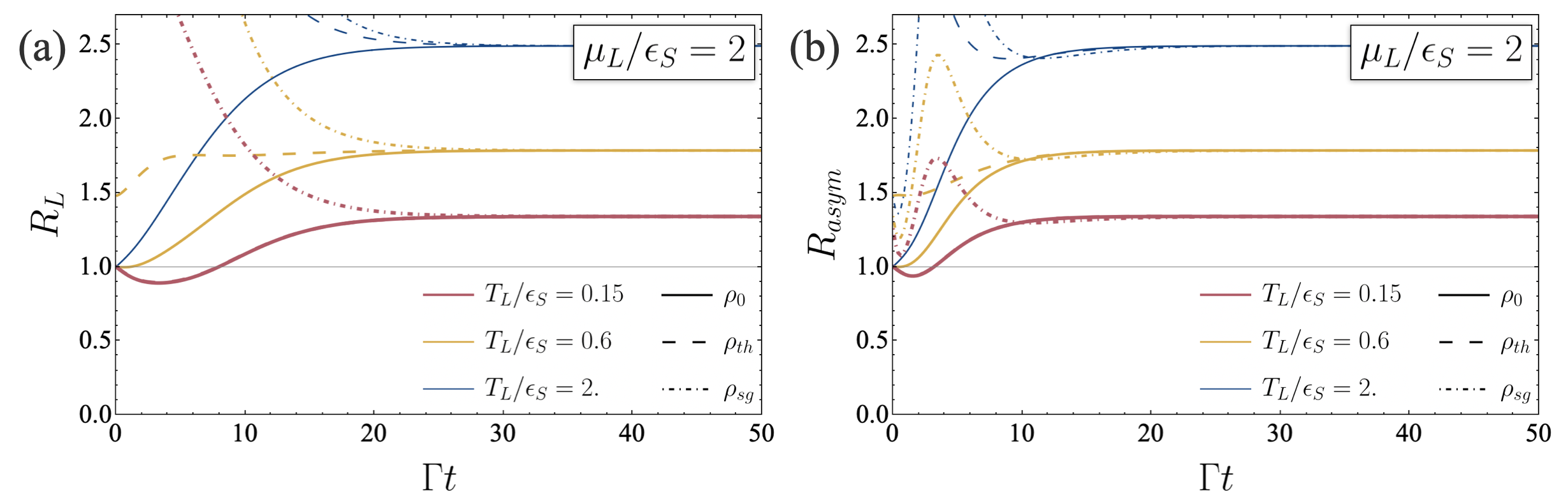}
\end{center}
\caption{``Left" (\textbf{a}) and ``asymmetric" (\textbf{b}) KUR ratios $R_L$ and $R_{asym}$ as functions of time, considering the activity $A = A_L + A_R + A_S$. Set of parameters: $\gamma_L/\epsilon_S=10^{-3}$, $\gamma_R/\epsilon_S=9\times10^{-3}$, $g/\epsilon_S=1.8\times10^{-3}$, $\mu_L/\epsilon_S=2$, $\mu_R/\epsilon_S=0$, $T_R/\epsilon_S = 0.1$.
\label{fig:KUR_fullA}}
\end{figure}

In Fig.~\ref{fig:KUR_fullA}, we plot the KUR ratios $R_L, R_{asym}$ accounting for this internal activity, \textit{i.e.} taking $A_{tot}(t) = A_L(t) + A_R(t) + A_S(t)$ in the definition of the KUR ratios.
As discussed in \cite{Prech2023}, internal activity can prevent KUR to be violated. This is the case in the steady-state for the set of parameters we choose for our plots. In the transient regime however, at low temperatures, KUR is still violated over a short time range for the small temperature $T_L=0.15 \epsilon_S$ and the ground initial state $\densmat_0$ (red solid line). We can not evidence a specific role of the presence of quantum correlations in this figure.


\end{document}

%% file: commands.tex
\newcommand{\I}{\mathbb{I}}

\newcommand{\densmat}{\rho}

\newcommand{\avg}[1]{\overline{#1}}

\newcommand{\gL}{\gamma_L}
\newcommand{\gR}{\gamma_R}
\newcommand{\gLp}{\gamma_L^+}
\newcommand{\gLm}{\gamma_L^-}
\newcommand{\gRp}{\gamma_R^+}
\newcommand{\gRm}{\gamma_R^-}

%% file: NJP_submission.bbl
\providecommand{\newblock}{}
\begin{thebibliography}{10}
\expandafter\ifx\csname url\endcsname\relax
  \def\url#1{{\tt #1}}\fi
\expandafter\ifx\csname urlprefix\endcsname\relax\def\urlprefix{URL }\fi
\providecommand{\eprint}[2][]{\url{#2}}

\bibitem{Kosloff2014}
Kosloff R and Levy A 2014 {\em Annual Review of Physical Chemistry\/} {\bf 65}
  365–393 ISSN 1545-1593
  \urlprefix\url{http://dx.doi.org/10.1146/annurev-physchem-040513-103724}

\bibitem{Vinjanampathy2016}
Vinjanampathy S and Anders J 2016 {\em Contemp. Phys.\/} {\bf 57}
  \urlprefix\url{https://www.tandfonline.com/doi/full/10.1080/00107514.2016.1201896}

\bibitem{Mitchison2019}
Mitchison M~T 2019 {\em Contemporary Physics\/} {\bf 60} 164–187 ISSN
  1366-5812 \urlprefix\url{http://dx.doi.org/10.1080/00107514.2019.1631555}

\bibitem{Bhattacharya2021}
Bhattacharjee S and Dutta A 2021 {\em Eur. Phys. J. B\/} {\bf 94} 239
  \urlprefix\url{https://doi.org/10.1140/epjb/s10051-021-00235-3}

\bibitem{Myers2022}
Myers N~M, Abah O and Deffner S 2022 {\em AVS Quantum Science\/} {\bf 4} ISSN
  2639-0213 \urlprefix\url{http://dx.doi.org/10.1116/5.0083192}

\bibitem{Eisler2005}
Eisler V and Zimborás Z 2005 {\em Physical Review A\/} {\bf 71} ISSN 1094-1622
  \urlprefix\url{http://dx.doi.org/10.1103/PhysRevA.71.042318}

\bibitem{Hartmann2007}
Hartmann L, D\"{u}r W and Briegel H~J 2007 {\em New Journal of Physics\/} {\bf
  9} 230–230 ISSN 1367-2630
  \urlprefix\url{http://dx.doi.org/10.1088/1367-2630/9/7/230}

\bibitem{Quiroga2007}
Quiroga L, Rodríguez F~J, Ramírez M~E and París R 2007 {\em Physical Review
  A\/} {\bf 75} ISSN 1094-1622
  \urlprefix\url{http://dx.doi.org/10.1103/PhysRevA.75.032308}

\bibitem{Bellomo2013}
Bellomo B and Antezza M 2013 {\em New Journal of Physics\/} {\bf 15} 113052
  ISSN 1367-2630
  \urlprefix\url{http://dx.doi.org/10.1088/1367-2630/15/11/113052}

\bibitem{BohrBrask2015}
Brask J~B, Haack G, Brunner N and Huber M 2015 {\em New Journal of Physics\/}
  {\bf 17} 113029
  \urlprefix\url{https://doi.org/10.1088/1367-2630/17/11/113029}

\bibitem{Hegwill2018}
Hewgill A, Ferraro A and De~Chiara G 2018 {\em Phys. Rev. A\/} {\bf 98}(4)
  042102 \urlprefix\url{https://link.aps.org/doi/10.1103/PhysRevA.98.042102}

\bibitem{Man2019}
Man Z~X, Tavakoli A, Brask J~B and Xia Y~J 2019 {\em Phys. Scr.\/} {\bf 94}
  075101 \urlprefix\url{https://dx.doi.org/10.1088/1402-4896/ab0c51}

\bibitem{Farina23}
Farina D, Benazout B, Centrone F and Acin A 2023
  \urlprefix\url{https://arxiv.org/abs/2307.13341}

\bibitem{Tavakoli2018}
Tavakoli A, Haack G, Huber M, Brunner N and Brask J~B 2018 {\em {Quantum}\/}
  {\bf 2} 73 ISSN 2521-327X
  \urlprefix\url{https://doi.org/10.22331/q-2018-06-13-73}

\bibitem{Tavakoli2020}
Tavakoli A, Haack G, Brunner N and Brask J~B 2020 {\em Phys. Rev. A\/} {\bf
  101}(1) 012315
  \urlprefix\url{https://link.aps.org/doi/10.1103/PhysRevA.101.012315}

\bibitem{Khandelwal2024}
Khandelwal S, Annby-Andersson B, Diotallevi G~F, Wacker A and Tavakoli A 2024
  {\em arXiv preprint arXiv:2401.01776\/}

\bibitem{Khandelwal2020}
Khandelwal S, Palazzo N, Brunner N and Haack G 2020 {\em New Journal of
  Physics\/} {\bf 22} 073039
  \urlprefix\url{https://iopscience.iop.org/article/10.1088/1367-2630/ab9983/meta}

\bibitem{BohrBrask2022}
Bohr~Brask J, Clivaz F, Haack G and Tavakoli A 2022 {\em {Quantum}\/} {\bf 6}
  672 ISSN 2521-327X \urlprefix\url{https://doi.org/10.22331/q-2022-03-22-672}

\bibitem{Prech2023}
Prech K, Johansson P, Nyholm E, Landi G~T, Verdozzi C, Samuelsson P and Potts
  P~P 2023 {\em Phys. Rev. Res.\/} {\bf 5}(2) 023155
  \urlprefix\url{https://link.aps.org/doi/10.1103/PhysRevResearch.5.023155}

\bibitem{Das2022}
Das A, Khan A~A, Mishra S~D, Solanki P, De B, Muralidharan B and Vinjanampathy
  S 2022 {\em Quantum Science and Technology\/} {\bf 7} 045034 ISSN 2058-9565
  \urlprefix\url{http://dx.doi.org/10.1088/2058-9565/ac8fb0}

\bibitem{Khandelwal2023}
Khandelwal S, Chen W, Murch K~W and Haack G 2023
  \urlprefix\url{https://arxiv.org/abs/2310.11381}

\bibitem{Tacchino2018}
Tacchino F, Auffèves A, Santos M and Gerace D 2018 {\em Physical Review
  Letters\/} {\bf 120} ISSN 1079-7114
  \urlprefix\url{http://dx.doi.org/10.1103/PhysRevLett.120.063604}

\bibitem{Heineken2021}
Heineken D, Beyer K, Luoma K and Strunz W~T 2021 {\em Phys. Rev. A\/} {\bf
  104}(5) 052426
  \urlprefix\url{https://link.aps.org/doi/10.1103/PhysRevA.104.052426}

\bibitem{Aguilar2020}
Aguilar M, Freitas N and Paz J~P 2020 {\em Physical Review A\/} {\bf 102} ISSN
  2469-9934 \urlprefix\url{http://dx.doi.org/10.1103/PhysRevA.102.062422}

\bibitem{Diotallevi2023}
Diotallevi G~F, Annby-Andersson B, Samuelsson P, Tavakoli A and Bakhshinezhad P
  2023 \urlprefix\url{https://arxiv.org/abs/2309.07696}

\bibitem{Blasi2023}
Blasi G, Khandelwal S and Haack G 2023 {\em arXiv preprint arXiv:2312.15065\/}
  \urlprefix\url{https://arxiv.org/abs/2312.15065}

\bibitem{Garrahan2017}
Garrahan J~P 2017 {\em Phys. Rev. E\/} {\bf 95}(3) 032134
  \urlprefix\url{https://link.aps.org/doi/10.1103/PhysRevE.95.032134}

\bibitem{Terlizzi2019}
Terlizzi I~D and Baiesi M 2018 {\em J. Phys. A: Math. Theor.\/} {\bf 52} 02LT03
  \urlprefix\url{https://dx.doi.org/10.1088/1751-8121/aaee34}

\bibitem{Hiura2021}
Hiura K and Sasa S~i 2021 {\em Physical Review E\/} {\bf 103} ISSN 2470-0053
  \urlprefix\url{http://dx.doi.org/10.1103/PhysRevE.103.L050103}

\bibitem{Potts2021}
Potts P~P, Kalaee A~A~S and Wacker A 2021 {\em New J. Phys.\/} {\bf 23} 123013
  \urlprefix\url{https://dx.doi.org/10.1088/1367-2630/ac3b2f}

\bibitem{Breuer2007}
Breuer H~P and Petruccione F 2007 {\em {The Theory of Open Quantum Systems}\/}
  vol~1 (Oxford University Press) ISBN 9780191706349

\bibitem{Schaller2014}
Schaller G 2014 {\em Dynamics of Open Quantum Systems\/} (Springer
  International Publishing) p 1–26 ISBN 9783319038773
  \urlprefix\url{http://dx.doi.org/10.1007/978-3-319-03877-3_1}

\bibitem{Hofer2017}
Hofer P~P, Perarnau-Llobet M, Miranda L~D~M, Haack G, Silva R, Brask J~B and
  Brunner N 2017 {\em New J. Phys.\/} {\bf 19} 123037 ISSN 1367-2630
  \urlprefix\url{https://iopscience.iop.org/article/10.1088/1367-2630/aa964f}

\bibitem{Baldea2016}
B\^aldea I 2016 {\em Beilstein Journal of Nanotechnology\/} {\bf 7} 418–431
  ISSN 2190-4286 \urlprefix\url{http://dx.doi.org/10.3762/bjnano.7.37}

\bibitem{Covito2018}
Covito F, Eich F, Tuovinen R, Sentef M and Rubio A 2018 {\em Journal of
  chemical theory and computation\/} {\bf 14} 2495--2504

\bibitem{Landi2023}
Landi G~T, Kewming M~J, Mitchison M~T and Potts P~P 2024 {\em PRX Quantum\/}
  {\bf 5}(2) 020201
  \urlprefix\url{https://link.aps.org/doi/10.1103/PRXQuantum.5.020201}

\bibitem{Wiseman}
Wiseman H~M and Milburn G~J 2009 {\em Quantum Measurement and Control\/}
  (Cambridge University Press)

\bibitem{Khandelwal2021}
Khandelwal S, Brunner N and Haack G 2021 {\em PRX Quantum\/} {\bf 2}(4) 040346
  \urlprefix\url{https://link.aps.org/doi/10.1103/PRXQuantum.2.040346}

\bibitem{Wootters1998}
Wootters W~K 1998 {\em Physical Review Letters\/} {\bf 80} 2245–2248 ISSN
  1079-7114 \urlprefix\url{http://dx.doi.org/10.1103/PhysRevLett.80.2245}

\bibitem{Yu2007}
Yu T and Eberly J 2007 {\em Quantum Information and Computation\/} {\bf 7}
  459–468 ISSN 1533-7146
  \urlprefix\url{http://dx.doi.org/10.26421/QIC7.5-6-3}

\bibitem{Timpanaro2019}
Timpanaro A~M, Guarnieri G, Goold J and Landi G~T 2019 {\em Physical Review
  Letters\/} {\bf 123} ISSN 1079-7114
  \urlprefix\url{http://dx.doi.org/10.1103/PhysRevLett.123.090604}

\bibitem{Kewming2023}
Kewming M~J, Kiely A, Campbell S and Landi G~T 2023
  \urlprefix\url{https://arxiv.org/abs/2308.07810}

\bibitem{Timpanaro2023}
Timpanaro A~M, Guarnieri G and Landi G~T 2023 {\em Physical Review B\/} {\bf
  107} ISSN 2469-9969
  \urlprefix\url{http://dx.doi.org/10.1103/PhysRevB.107.115432}

\bibitem{Liu2021}
Liu J and Segal D 2021 {\em Physical Review E\/} {\bf 103} ISSN 2470-0053
  \urlprefix\url{http://dx.doi.org/10.1103/PhysRevE.103.032138}

\bibitem{Gerry2022}
Gerry M and Segal D 2022 {\em Physical Review B\/} {\bf 105} ISSN 2469-9969
  \urlprefix\url{http://dx.doi.org/10.1103/PhysRevB.105.155401}

\bibitem{Ferri2023}
Ferri-Cortés M, Almanza-Marrero J~A, López R, Zambrini R and Manzano G 2023
  \urlprefix\url{https://arxiv.org/abs/2308.08491}

\bibitem{Joho2012}
Joho K, Maier S and Komnik A 2012 {\em Phys. Rev. B\/} {\bf 86}(15) 155304
  \urlprefix\url{https://link.aps.org/doi/10.1103/PhysRevB.86.155304}

\bibitem{VanVu2022}
Van~Vu T and Saito K 2022 {\em Physical Review Letters\/} {\bf 128} ISSN
  1079-7114 \urlprefix\url{http://dx.doi.org/10.1103/PhysRevLett.128.140602}

\end{thebibliography}
